\documentclass{aa}
\usepackage{graphicx}
\begin{document}
   	\title{A search for planetary-mass objects and brown dwarfs in the Upper
	Scorpius association}


   	\author{M. T. Costado\inst{1}
		\and
   	  	V. J. S. B\'ejar\inst{2}
          	\and
		J. A. Caballero\inst{1}
          	\and
   	  	R. Rebolo\inst{1,3}
          	\and
		J. Acosta-Pulido\inst{1}
          	\and
   	  	A. Manchado\inst{1,3}
		}

   	\institute{Instituto de Astrof\'{\i}sica de Canarias, E-38205 La
	Laguna, Tenerife, Spain\\
        \email{vbejar@iac.es}
        \and
        Instituto de Astrof\'{\i}sica de Canarias. Gran Telescopio Canarias Project
        \and
        Consejo Superior de Investigaciones Cient\'{\i}ficas, Spain}

   	\date{Received March 11, 2005; accepted August 03, 2005}

   	\abstract{We report the results of a deep photometric search for planets
	and brown dwarfs in the nearby young OB Upper Scorpius association. 
	We obtained optical ($I$) and near-infrared ($JK_{\rm s}$) images around
	nine very low-mass stars and brown dwarf member candidates of the
	association, covering a total area of 113\,arcmin$^2$. 
	Using a point spread function subtraction technique, we have searched
	for planetary-mass companions (0.002\,$M_\odot$ $<$ M $<$ 0.013\,$M_\odot$) at
	separations as close as 3\arcsec\, from the targets.
	We have not found any brown dwarfs more massive than 0.030\,$M_\odot$ at
	projected distances larger than 70\,AU, or planets more massive than
	0.007\,$M_\odot$ at projected distances larger than 600\,AU. 
	We set an upper limit of 20\% (confidence level 68.3\,\%) to the
	fraction of very low-mass stars and massive brown dwarfs with planetary
	mass companions (M $>$ 0.007\,$M_\odot$) at physical distances larger
	than 600\,AU.  
	From the $I$, $I-J$ colour-magnitude diagrams and follow-up 
	$K_{\rm s}$-band photometry, we identify four very red objects ($I-J >$
	2.5) in the area of the survey.
	According to their positions in the diagram, below the expected
	theoretical sequence of the cluster, three are probable late M-dwarfs in
	the field.
	The faintest and reddest object ($J-K_{\rm s} =$ 2.39 $\pm$ 0.06) may be
	either a field intermediate L-dwarf with anomalously red $J-K_{\rm s}$
	colour at a distance of $\sim$135\,pc, an extremely red distant galaxy,
	or a reddened planetary-mass object in the Upper Scorpius association.  
	}
	
	\titlerunning{Planetary-mass objects in Upper Scorpius} 

   	\maketitle
%

\section{Introduction}

The Upper Scorpius OB association, located at a distance of 145 $\pm$ 2\,pc 
(de Zeeuw et al. \cite{dez99}) and with an estimated age of 5--6\,Myr 
(Preibisch \& Zinnecker \cite{pre99}), belongs to the 
Scorpio--Centaurus--Lupus--Crux complex.
The galactic latitude is in the range $+$19--21$^\circ$. 
The characteristics of youth, proximity and low extinction makes Upper Scorpius
an ideal site to search for very low-mass stars and substellar objects. 
Recent studies by Ardila et al. (\cite{ard00}) and Preibisch et al.
(\cite{pre01}, \cite{pre02}) have investigated the population of very low-mass
stars and brown dwarfs in the Upper Scorpius.  
Many of them have been confirmed as bona fide members of the association and
several show evidence of the presence of accretion discs (Jayawardhana
et al. \cite{jaya02}, \cite{jaya03}). 

In this paper we present deep images in the optical $I$ and infrared $JK_{\rm
s}$-bands of low-mass stars and brown dwarfs in the Upper Scorpius association
with sufficient sensitivity to directly detect planetary-mass companions at wide
separations.

\section{Sample selection and observations}

We selected nine faint and cool potential members of the Upper Scorpius
association from studies by Ardila et al. (\cite{ard00}) and Preibisch et al.
 (\cite{pre01}).  
The four objects selected from Preibisch et al. (\cite{pre01}) are very low-mass stars
of spectral type M\,5, with lithium in absorption and H$\alpha$ in emission, so
they are very likely members of this association.
Three out of the five objects selected from Ardila et al. (\cite{ard00}) were
 later confirmed as bona fide members through the detection of lithium in
their spectra.
They are very low-mass stars or massive brown dwarfs with
spectral types later than M\,5.  
They also display $L'$-band excess and/or very intense and
broad H$\alpha$ emission lines that suggest the existence of discs.  
Subsequent studies to our observations have shown that one of the remaining candidates
(UScoCTIO 121) has depleted its lithium (Muzerolle et al. \cite{muz03}) 
 and is probably a background field dwarf star.
Names, coordinates, 
photometric and main spectral properties for these
objects are given in Tables \ref{photo} and \ref{spec}
 (both tables are only available 
in the electronic version in Appendix \ref{tablas}).

We have performed optical ($I$) and near-infrared ($JK_{\rm s}$) observations. 
A log of the observations is presented in Table \ref{obser}. Standard reduction 
processes and aperture and psf photometry have been performed. 
For the calibration of the $J$ and $K_{\rm s}$-band, we used the photometry from 
the 2MASS catalogue (Cutri et al. \cite{cutri03}) and for the $I$-band 
data we used the photometry from the DENIS catalogue (DENIS Consortium 2003, 
 Epchtein et al. \cite{ed1994}).
We have tranformed the $I$-band data from DENIS into the Cousins system, 
using $I_{\rm C}-i_{\rm DENIS} = $ 0.030 $\pm$ 0.04, obtained by comparison of the photometry of M
and L dwarfs from Legget et al. (\cite{leg00}, \cite{leg01}) and DENIS.
Estimates for the completeness and limiting magnitudes (with an error of 0.5 mag) 
of the $I$-, $J$-, $K_{\rm s}$({\sc tcs})- and  $K_{\rm s}$({\sc liris})-band images are 
21.6, 19.2, 17.1, 18.5 and 22.5, 20.0, 17.8, 19.5 respectively.  
For more details see the 
electronic version in Appendix \ref{obs}. 

\begin{table*}
    \centering
    	\caption[]{Log of the observations}  
        \label{obser}
        \tiny{
	\begin{tabular}{lcccccccc}
            	\hline
            	\noalign{\smallskip}
Filter & Instrument & Telescope & Detector & Scale & Bin & Seeing & Date \\
\hline
 $I$ & OIG & TNG & 2048$\times$4096 EEV & 0.072\,arcsec/pixel & 2$\times$2 & 0.6-0.9 & 02Jun28-30 \\
 $J$ & INGRID & WHT & 1024 $\times$ 1024 pixel$^2$ Hawai'i & 0.238\,arcsec/pixel & 1$\times$1 & 1.1 & 02March26 \\
 $K_{\rm s}$ & CAIN-2 & TCS & 256 $\times$ 256 Nicmos-3 & 1\,arcsec/pixel & 1$\times$1 & 1.5 & 02June7-9 \\
 $K_{\rm s}$ & LIRIS & WHT & 1024 $\times$ 1024 pixel$^2$ Hawai'i & 0.25\,arcsec/pixel & 1$\times$1 & 1.2-1.5 & 04June8-10 \\
	\noalign{\smallskip}
    	\hline
        \end{tabular}
	}
\end{table*}

\section{Results and discussion}

\subsection{Search for nearby companions}

   \begin{figure}
   \centering
\includegraphics[width=0.15\textwidth]{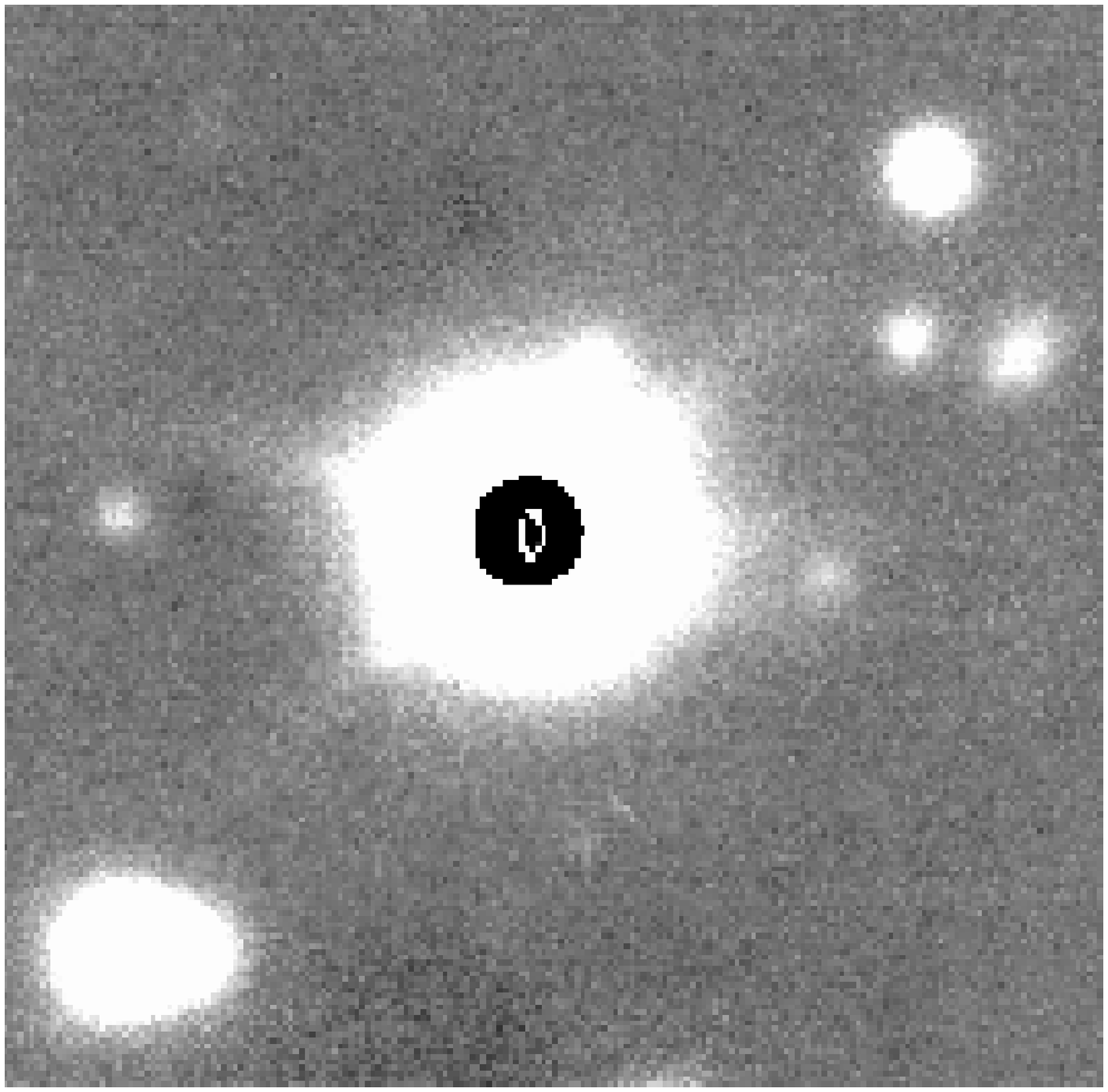}
\includegraphics[width=0.15\textwidth]{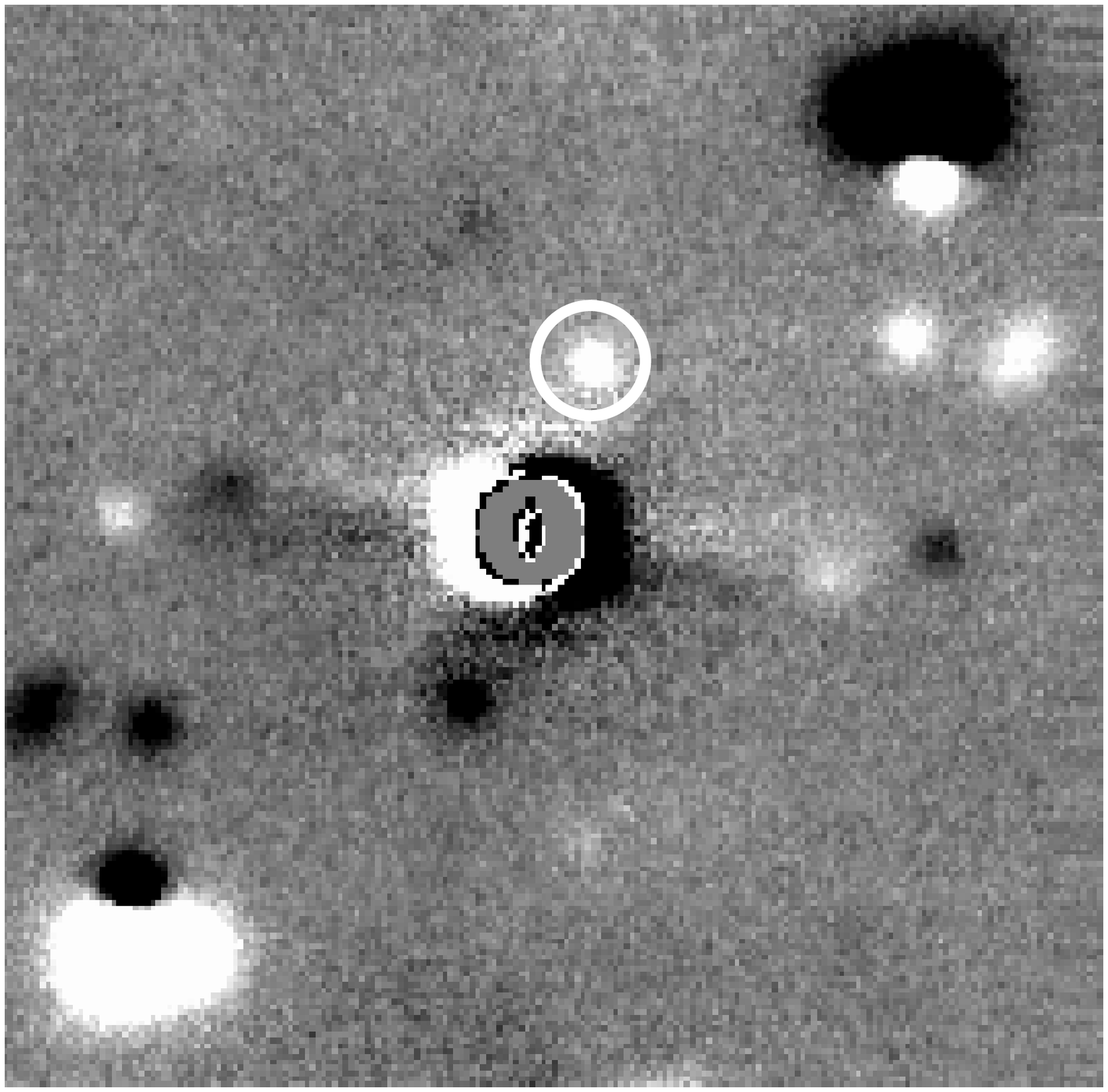}
\includegraphics[width=0.15\textwidth]{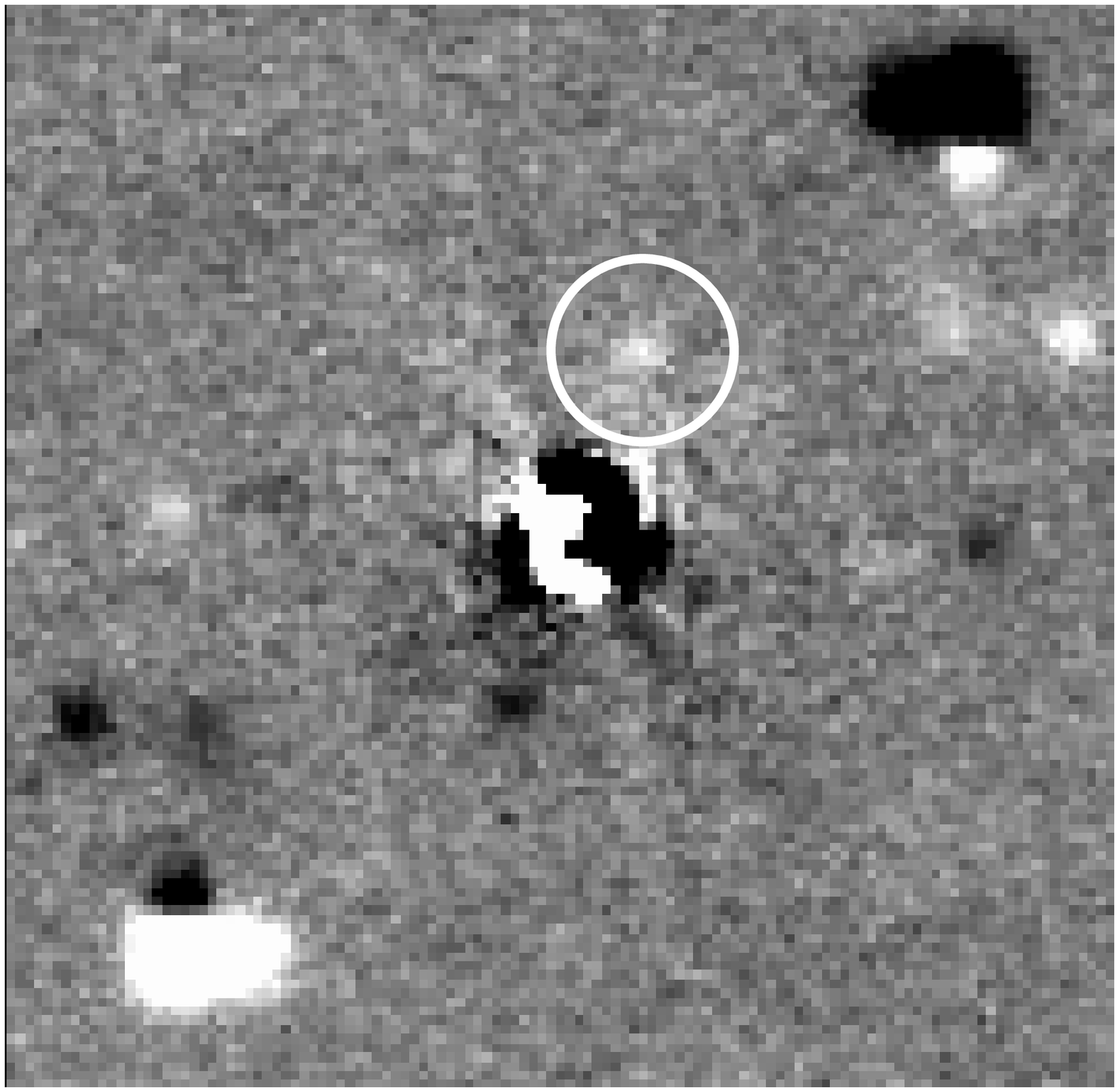}
      \caption{$I$- and $J$-band images of USco J160628.7$-$200357 and its
      nearby visual companion, marked with a white circle. 
      We show the $I$-band images prior to PSF subtraction in the left window.
      PSF-subtracted $I$- and $J$-band images are shown in the central and 
      right windows, respectively.
      The size of the images are 30$\times$30\,arcmin$^2$.
      North up, and east is to left.}
         \label{neighbours}
   \end{figure}


\begin{table*}
    \centering
    	\caption[]{Detected visual companions to our targets}  
        \label{companions}
        \tiny{
	\begin{tabular}{lcccccc}
            	\hline
            	\noalign{\smallskip}
USco primary 		& $\rho$  	&  $\theta$    		& $I_{\rm obj}$	& $(I-J)_{\rm obj}$ 	& $(I-J)_{\rm USco}$	& $I_{\rm obj} - I_{\rm USco}$ 	 \\
			& ($''$)  	&   (deg)      		&			&		 	&			& \\
\noalign{\smallskip}
\hline
\noalign{\smallskip}
UScoCTIO 56 		& 3.00$\pm$0.05	& 201.3$\pm$0.6          & 20.0$\pm$0.4       & 0.5$\pm$0.4	    & 2.07$\pm$0.05	    & 6.0$\pm$0.4 \\
USco J160628.7$-$200357	& 4.80$\pm$0.05	& 339.9$\pm$0.5          & 20.8$\pm$0.4       & 1.7$\pm$0.4	    & 1.85$\pm$0.06	    & 7.5$\pm$0.4 \\
USco J160629.0$-$205216	& 5.10$\pm$0.05	& 354.4$\pm$0.3          & 21.3$\pm$0.4       & 1.8$\pm$0.4	    & 2.20$\pm$0.05	    & 8.0$\pm$0.4 \\
           	\noalign{\smallskip}
            	\hline
         \end{tabular}
	 }
\end{table*}
 
We have performed a search for faint companions at close angular distances,
$\rho <$ 10\,arcsec, to the selected targets 
(equivalent to a projected distance $r <$ 1500\,AU).
Because of the depth of our images, the primaries are very close to saturation
and their wings are very strong and extended.  
To detect possible faint companions, we have implemented a PSF subtraction
method consisting
basically of the subtraction to the original image of the rotated
image centred on the primary.   
We  applied a rotation of 180$^\circ$ because this rotation reflects 
 the symmetry of the PSF quite well and determined the centre 
 of rotation by minimizing the resulting residuals 
of the subtraction of the wings of the central object. 
We have found only three possible candidates located at separations of 3 to
5\,arcsec from three primaries (see Figure \ref{neighbours} and Table
\ref{companions}).    
These candidates are detected both in the $I$ and $J$ images, but not in the
$K_{\rm s}$ images because of the lower resolution of the images taken with 
{\sc cain-2}.  
   
We have performed aperture photometry of these candidates in the PSF subtracted
image. 
 The magnitude $I$ and the colour $(I-J)$ of the close visual
companions (obj), of their primaries (USco) and the difference in magnitude of both
 are shown in Table \ref{companions}.
Although uncertainties in the $IJ$-band magnitudes of the visual companions
are very large, the candidates are bluer than the primary central objects, 
opposite to what is expected if they were real companions. 
Hence they seem to be just field objects.

\subsection{Detectability of nearby companions}

We have estimated the faintest magnitude (i.e. lowest mass) that can be detected
at a given distance of a target in our images. 
We have generated artificial objects with different 
magnitudes using the {\tt mkobjects} routine within {\sc iraf}. 
After PSF subtraction, using the method explained in the previous section,
we determined the magnitude of the faintest object that can be detected by
visual inspection. 
This analysis was repeated for a range of distances to the central object 
until a distance where there is no major contribution from the primary wings and 
the sources are detected at the limiting magnitude of the survey. 
 
\begin{figure}
   \centering
\includegraphics[width=0.49\textwidth]{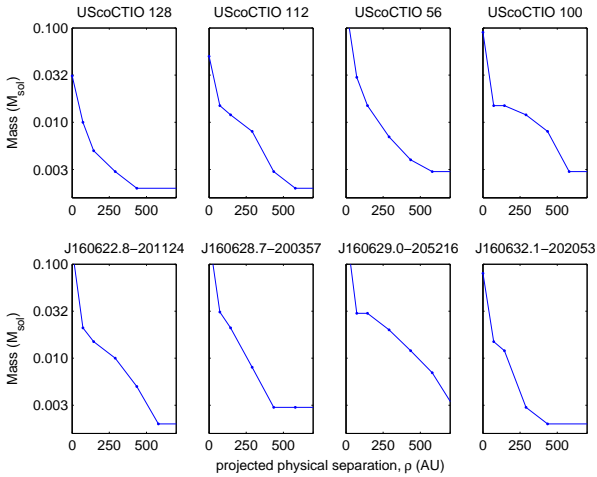}
      \caption{The minimum mass of detectable companions at a given
      distance of the targets. 
      The vertical scale is logarithmic.
}
         \label{simulados}
   \end{figure}

We have transformed $J$-band magnitude to masses, using the theoretical isochrone of 5\,Myr from
the Lyon group (Baraffe et al. \cite{bar98}; Chabrier et al. \cite{cha00}), and a
distance modulus $J-M_{J} =$ 5.81 $\pm$ 0.03 for the Upper Scorpius association (de
Zeeuw et al. \cite{dez99}). 
We have converted the luminosity and $T_{\rm eff}$ given by the theoretical 
models into magnitudes using the bolometric corrections and $T_{\rm eff}$ vs. spectral type from 
Golimowski et al. (\cite{gol04}), 
and the spectral type vs. $I-J$ colour relation from Leggett et al. 
(\cite{leg00}, \cite{leg01}). 
In Figure \ref{simulados}, we show the minimum masses of the
companions that can be detected 
at angular separations between 0.5\,arcsec ($\sim$70\,AU) and 5\,arcsec (750\,AU). 
The mass of the central object is included at the null distance. 
We have the sensitivity to detect planets more 
massive than 0.002\,$M_\odot$ ($\sim$2\,$M_{\rm Jup}$) 
and brown dwarfs
at angular separations larger than 3--4\,arcsec (450--600\,AU)
and 0.5--3\,arcsec ($\sim$70--450\,AU), respectively.
In conclusion, we have not found any brown dwarfs more massive than
0.030\,$M_\odot$ at projected distances larger than 70\,AU, or any planets more
massive than 0.007\,$M_\odot$ at projected distances larger than 600\,AU.

As we have not found any physical companion, 
we can only determine an upper limit of 14.3\,\% (for a confident level of 68.3\,\%)
 to the real fraction of substellar companions at the distances and masses mentioned above
 (see the Appendix \ref{estadistica} 
in electronic version for more details). 
In order to know the real fraction of substellar companions at physical
(not projected) distances, it is necessary to correct this result by the
probability that the orbital position of the companion is at a detectable
projected distance. 
Taking into account this correction,
we can determine an upper limit of 20\,\% on the
fraction of very low-mass stars and brown dwarfs with planetary companions ($M
>$ 0.007\,$M_\odot$) at physical distances larger than $\sim$600\,AU and with
brown dwarf companions at distances larger than $\sim$400\,AU. 
Less restrictive upper limits can be set for shorter distances. 
These results together with McCarthy \& Zuckerman
(\cite{mccarthy04}) seem to indicate that substellar objects are rare
companions of low-mass stars. 
The discovery of a planet candidate by Chauvin et al. (\cite{cha04}), however,
shows that the probability of finding planets around substellar objects could
be low, but not null.  

\subsection{Search for isolated low mass members}
\label{isolated}

\begin{table*}
    \centering
    	\caption[]{Ultracool objects}  
        \label{candidates}
        \tiny{
	\begin{tabular}{lcccccccc}
            	\hline
            	\noalign{\smallskip}
Id. 	&	$\alpha$	& $\delta$ 	  & $I$			  & $I-J$		  & $J-K_{\rm s}$ 	  & Expected 	& Field \\
number	&	(J2000)		& (J2000)	  & 		  	  & 			  &  			  & Sp.T.$^a$ 	&  \\
\noalign{\smallskip}
\hline
\noalign{\smallskip}
\#1	&	16 06 26.0	& $-$20 11 52	  & 22.5 $\pm$ 0.5	  & 2.9 $\pm$ 0.5	  & $<$ 2.3		  & M\,7 to L\,6  & USco J160622.8$-$201124 \\
\#2	&	16 06 25.2	& $-$20 04 41	  & 22.22 $\pm$ 0.09	  & 3.04 $\pm$ 0.09	  & $<$ 1.9		  & M\,8:  	& USco J160628.7$-$200357 \\
\#3	&	16 06 31.2	& $-$20 53 50	  & 21.43 $\pm$ 0.08	  & 3.07 $\pm$ 0.09	  & 1.01 $\pm$ 0.09	  & M\,8:  	& USco J160629.0$-$205216 \\
\#4	&	16 06 29.1	& $-$20 19 28	  & 23.0 $\pm$ 0.5	  & 3.6 $\pm$ 0.5	  & 2.39 $\pm$ 0.1	  & L\,5--L\,7:  & USco J160632.1$-$202053 \\
           	\noalign{\smallskip}
            	\hline
         \end{tabular}
	\begin{list}{}{}
	\item $^a$ Estimated spectral types assuming that the objects are dwarfs. 	
	\end{list}	
	}
\end{table*}
 
\begin{figure}
   \centering
\includegraphics[width=0.49\textwidth]{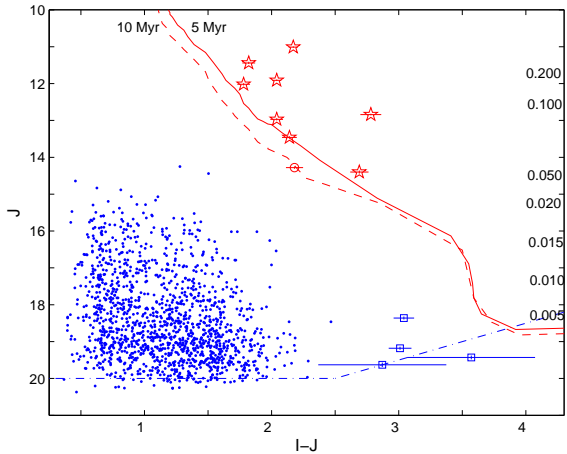}
      \caption{$I, I-J$ Colour-Magnitude diagram. 
      Previously known Upper Scorpius members are indicated with (red) stars. 
      Probable non-member UScoCTIO 121 is marked with a (red) circle.
      (Blue) squares denotes very cool objects found in our search and field
      stars are represented by (blue) dots. 
      Solid and dashed (red) lines correspond to the 5 and 10\,Myr theoretical
      isochrones from the Lyon group respectively (Baraffe et al. \cite{bar98};
      Chabrier et al. \cite{cha00}).
      Limit magnitude is marked with a (blue) dot-dashed line.
      Theoretical masses ($M_\odot$) for the 5-Myr Lyon model, are shown to
      the right.}
         \label{dcmg}
   \end{figure}

We have also conducted a conventional search for free-floating substellar
objects in our images building a $J$, $I-J$ colour--magnitude diagram for each
field.   
In Figure \ref{dcmg} we show the combination of all of them. 
Limiting magnitude is represented by a dot-dashed line. 
The 5 and 10\,Myr isochrones from evolutionary models of the Lyon group  
(Baraffe et al. \cite{bar98}; Chabrier et al. \cite{cha00}) are also plotted. 
We have converted their luminosity and $T_{\rm eff}$ 
into $I-J$ colours and $J$ magnitudes following the prescriptions given before. 
From this diagram, we have selected the reddest object ($I-J \ge$
2.5) identified in this survey. The photometry of these four ultracool 
objects are shown in Table \ref{candidates}. Since the optical 
and infrared images were not obtained simultaneously, the $I-J$ colour of these objects 
could be affected by intrinsic variability if they are really young objects. 

From their $I-J$ and $J-K_{\rm s}$ colours, 
we determined that objects with identification numbers \#2 and \#3 correspond to dwarfs
with M8 spectral type and that object \#1 has a spectral
type later than M\,7--M\,8 and earlier than L\,6. 
The last object, identified as \#4, requires especial attention.
According to the extremely red $I-J$ colour, the object is
expected to have a spectral type at around L\,5--L\,7 with a very red $J-K_{\rm s}$ 
colour (2.39 $\pm$ 0.06). 
If the object is a field L\,6 dwarf, it would be
a brown dwarf with an estimated mass of 0.060--0.072\,$M_\odot$ (using Dusty00
 models by Chabrier et al. 2000)
located at a distance of  around 135\,pc.
We cannot either rule out the possibility that object \#4 is a 
modestly extincted ($A_V =$ 2--3) isolated planetary-mass object in Upper Scorpius
or an unresolved ({\sc fwhm} $<$ 1\arcsec) extremely red galaxy.
We estimate from searches for extremely red objects (Thompson et al. \cite{tho99}) that
less than one such unresolved object might contaminate our survey.

According to their position in the $J$, $I-J$ color-magnitude diagram, 
these objects show $J$ magnitude fainter and $I-J$ colours bluer than the 
10 Myr isochrone, which is the oldest age expected for the association, so they
do not probably belong to the association.
This conclusion is in agreement with the number of contaminants we estimate in
our survey  from the densities of field M, L and T dwarfs derived in Kirkpatrick
et al. (\cite{kirk94}), Cruz et al. (\cite{cruz03}) and Burgasser et al.
(\cite{bur03}), respectively.  
We expect that around 2 late M, 0.7 L and no T field dwarfs will be present
in our search given the $I$-band limiting magnitude and the total area of the
survey.

\section{Conclusions}

We have performed a search for substellar companions around eight 
very low-mass stars and brown dwarfs of the Upper Scorpius association. 
Using a PSF subtraction method, we have found three faint objects at
3--5\,arcsec of the targets, but according to their $I-J$ colours, they seem to
be background visual companions.  
In conclusion, we have not found any brown dwarfs more massive than
0.030\,$M_\odot$ at projected distances larger than 70\,AU, and planets more
massive than 0.007\,$M_\odot$ at projected distances larger than 600\,AU. 
From the negative results of our search, we infer that the fraction of 
very-low mass stars and massive brown dwarfs with planets more massive than 
$\sim$0.007\,$M_\odot$ at physical distances larger than 600\,AU is lower than
20\,\% (confidence level of 68.3\,\%).  
Using $I$, $I-J$ colour-magnitude diagrams and follow-up $K_{\rm s}$-band
photometry, we have identified four ultracool candidates. 
According to their positions in the $I$, $I-J$ diagram, below the expected 
theoretical sequence of the cluster, three are probable late M/early L dwarfs in 
the field. The number of contaminants presented in our survey is consistent with our
estimates. 
The faintest and reddest object ($J-K_{\rm s} =$ 2.39 $\pm$ 0.06) could be 
a field intermediate L dwarf with anomalously red $J-K_{\rm s}$ 
colour at a distance of $\sim$135\,pc, an extremely red galaxy, or a reddened
planetary-mass object in the Upper Scorpius association. 
A proper motion determination and follow-up spectroscopy are required to
disentangle  these possibilities.

\begin{acknowledgements}
  	We would like to thank to M. Rosa Zapatero Osorio for providing valuable software 
	and Terry Mahoney for revising the English of the manuscript. 
	AM and JAP acknowledge support from grant AYA 2004-03136, financed
	by the Spanish Direcci\'on General de Investigaci\'on.
\end{acknowledgements}

\appendix

\section{Tables}
\label{tablas}
\begin{table*}
    \centering
    	\caption[]{Name, coordinates and $IJK_{\rm s}$ magnitudes of the targets}  
        \label{photo}
	\tiny{
        \begin{tabular}{lccccc}
            	\hline
            	\noalign{\smallskip}
Name 				& $\alpha$	& $\delta$	& $I$			& $J$  			& $K_{\rm s}$  	 	\\
				& (J2000.0)	& (J2000.0)	& (COUSINS)		& (2MASS) 		& (2MASS)      		\\
\noalign{\smallskip}
\hline
\noalign{\smallskip}
UScoCTIO 121$^a$		& 15 51 47.3	& $-$26 23 47	& 16.49$\pm$0.07	& 14.28$\pm$0.03	& 13.32$\pm$0.04	\\
UScoCTIO 128			& 15 59 11.2	& $-$23 37 59	& 17.12$\pm$0.07	& 14.40$\pm$0.04	& 13.21$\pm$0.04	\\
UScoCTIO 112			& 16 00 26.6	& $-$20 56 32	& 15.63$\pm$0.06	& 13.46$\pm$0.03	& 12.51$\pm$0.02	\\
UScoCTIO 56 			& 16 01 41.0	& $-$20 22 08	& 13.98$\pm$0.04	& 11.91$\pm$0.02	& 10.86$\pm$0.02	\\
UScoCTIO 100			& 16 02 04.1	& $-$20 50 42	& 15.65$\pm$0.09	& 12.84$\pm$0.02	& 11.83$\pm$0.03	\\
USco J160622.8$-$201124		& 16 06 22.8	& $-$20 11 24	& 13.83$\pm$0.05	& 12.02$\pm$0.03	& 11.002$\pm$0.019	\\
USco J160628.7$-$200357		& 16 06 28.7	& $-$20 03 57	& 13.29$\pm$0.05	& 11.44$\pm$0.03	& 10.464$\pm$0.019	\\
USco J160629.0$-$205216		& 16 06 29.0	& $-$20 52 16	& 13.21$\pm$0.04	& 11.01$\pm$0.02	& 10.00$\pm$0.02	\\
USco J160632.1$-$202053		& 16 06 32.1	& $-$20 20 53	& 15.04$\pm$0.05	& 12.97$\pm$0.03	& 11.94$\pm$0.02	\\
           	\noalign{\smallskip}
            	\hline
         \end{tabular}
	\begin{list}{}{}
	\item $^a$ Probable non-member.
	\end{list}
	}
\end{table*}

\begin{table*}
    \centering
    	\caption[]{Name, main spectral properties and 
	references of the targets}  
        \label{spec}
	\tiny{
        \begin{tabular}{lccccc}
            	\hline
            	\noalign{\smallskip}
Name 				& EW(Li)	& EW(H$\alpha$) & Spectral 	& Remarks$^a$	& References$^b$\\		 
				& (\AA)		& (\AA)		& type 		&		& \\			
\noalign{\smallskip}
\hline
\noalign{\smallskip}
UScoCTIO 121			& $\le$0.3	& $-$10.0	& M\,6 		& $Nm$		&  1, 6, 9 \\		
UScoCTIO 128			& 0.5		& $-$130.5	& M\,7 		& $L$, $lowg$, $Em$ & 1, 4, 5, 6, 7, 9 \\
UScoCTIO 112			& 0.6		& $-$16.1	& M\,5.5 		& $Em$		&  1, 4, 7, 9 \\ 	 
UScoCTIO 56 			& --		& $-$9.1	& M\,5 		& --		&  1, 2 \\		
UScoCTIO 100			& 0.6		& $-$16.1	& M\,7 		& $Em$		&  1, 4, 5, 6, 7, 8, 9 \\
USco J160622.8$-$201124		& 0.53		& $-$6.0	& M\,5 		& --		&  2, 3 \\		
USco J160628.7$-$200357		& 0.58		& $-$30.0	& M\,5 		& --		&  2, 3 \\		
USco J160629.0$-$205216		& 0.68		& $-$6.2	& M\,5 		& --		&  2, 3 \\		
USco J160632.1$-$202053		& 0.69		& $-$5.1	& M\,5 		& --		&  2, 3 \\		
           	\noalign{\smallskip}
            	\hline
         \end{tabular}
	\begin{list}{}{}
	\item $^a$
	Nm.: Probable non-member;
	$L$: Excess in the $L'$ band;
	$lowg$: Low surface gravity;
	$Em$: Presence of lines in emission;
	
	\item $^b$ References:
	(1) Ardila et al. \cite{ard00};
	(2) Preibisch et al. \cite{pre01};
	(3) Preibisch et al. \cite{pre02};
	(4) Jayawardhana et al. \cite{jaya02};
	(5) Gorlova et al. \cite{gor03};
	(6) Muzerolle et al. \cite{muz03};
	(7) Jayawardhana et al. \cite{jaya03};
	(8) G\"unther \& Wuchterl \cite{gun03};
	(9) Mohanty et al. \cite{moh04}.
	\end{list}	
	}
\end{table*}

\section{Observations and data analysis}
\label{obs}

\subsection{Optical observations}

We obtained Cousins $I$-band images using the Optical Imager Galileo
({\sc oig}) instrument, mounted on a Nasmyth platform of the Telescopio Nazionale
Galileo at the Roque de los Muchachos Observatory, La Palma, on 2002 June 28--30. 
This optical camera has two 2048 $\times$ 4096 pixel EEV 42-80 CCD
detectors, with a plate scale of 0.072\,arcsec/pixel, and covering a total
area of 4.9 $\times$ 4.9 arcmin$^2$ in each exposure.  
There is a physical separation between both detectors of $\sim$2.8\,arcsec. 
Spatial binning of 2 $\times$ 2 pixels was used during our observations
(hence, giving a real pixel size of 0.144\,arcsec). 
Nights were photometric and average seeing ranged from 0.6 to 0.9\,arcsec.  
Raw data were bias-subtracted using the overscan region and flat-field
corrected. 
Flat-field images for each night were obtained combining sky flat-field images
taken at dusk.  
For the reduction we used routines from the {\tt imred} package within the
{\sc iraf}\footnote{IRAF is distributed by National Optical Astronomy
Observatories, which is operated by the Association of Universities for
Research in Astronomy, Inc., under contract with the National Science
Foundation.} environment.

\subsection{Infrared Observations}

We obtained $J$-band images using the {\sc ingrid} instrument, mounted on the
Cassegrain focus of the 4.2\,m William Herschel Telescope at the Roque de
los Muchachos Observatory, La Palma, on 2002 March 26.   
{\sc ingrid} is a near-infrared camera which consists of a 1024 $\times$ 1024
pixel$^2$ Hawai'i detector, providing a plate-scale of 0.238 arcsec/pixel, and
covering a total area of 4.06 $\times$ 4.06 arcmin$^2$ in each exposure.    
The night was photometric and the average seeing was of 1.1\,arcsec. 
Observations were performed using a dithering pattern of nine positions. 
For each field, the total exposure time was 9 $\times$ 4 $\times$ 30\,s =
1080\,s. 

$K_{\rm s}$-band images for all the fields except for USco J160632.1$-$202053
were obtained with the {\sc cain-2} instrument, mounted on the 1.55\,m
Telescopio Carlos S\'anchez at the Teide Observatory on 2002 June 7--9. 
The {\sc cain-2} near-infrared camera has a 256 $\times$ 256 Nicmos-3 detector,
and in its wide-optic mode provides a plate scale of 1.00\,arcsec/pixel and a
total area of 4.27 $\times$ 4.27 arcmin$^2$. 
The nights were not photometric (uniform layer of thin cirrus), that slightly 
affects the limiting magnitude of our observations, 
 and the average seeing was  1.5\,arcsec.  
We performed the observations using a dithering pattern of ten positions, with
6--12 exposures of 20--10\,s in each position, giving a total exposure time of
1800\,s.  
The $K_{\rm s}$-band image of the USco J160632.1$-$202053 and UScoCTIO 128 fields
were obtained with the {\sc liris} instrument mounted on the Cassegrain focus of
the 4.2\,m William Herschel Telescope on 2004 June 8 and 10, as part of the
Guaranteed Time programme. 
Thus, two different $K_{\rm s}$-band images for UScoCTIO 128 were obtained.  
The {\sc liris} instrument is a new near-infrared camera and spectrograph,
built at the Instituto de Astrof\'{\i}sica de Canarias (Acosta-Pulido et al.
\cite{ap03}; Manchado et al. \cite{ma04}). 
It uses a 1024 $\times$ 1024 Hawai'i detector, providing a plate scale of
0.250\,arcsec/pixel and a total area of 4.27 $\times$ 4.27 arcmin$^2$.   
The nights were photometric and the average seeing was  in the range of 1.2--1.5\,arcsec. 
A five-point dithering pattern was used, obtaining 15 frames of 8\,s exposure time at
each point, which means an exposure time per cycle of 600\,s. 
This cycle was repeated six times giving a total exposure time of 3600\,s.

All infrared data were processed in the same way. 
Raw data were sky-subtracted and flat-field corrected.
We generated sky frames for each field and filter combination by median-filtering the 
dithered scientific images to eliminate the point sources. Flat-field images were derived  
combining sky flat-field images taken at dawn and dusk for the {\sc ingrid} and {\sc
cain-2} data. 
Dome flat-field images were obtained for the {\sc liris} data. 
Images were then aligned and combined.
For the reduction we used several routines within the {\sc iraf} environment.

\subsection{Photometric analysis}

For all the images obtained in different filters we carried out the same
photometric analysis.
We used routines from the {\tt daophot} package within {\sc iraf}. 
Point sources were selected using the {\tt daofind} task, so extended sources were 
mostly avoided. We performed aperture and psf photometry of the selected objects using
the {\tt phot} routine. 
To tranform instrumental into real magnitudes we used the photometry of objects in common 
with our images and several existing catalogues.
For the calibration of the $J$ and $K_{\rm s}$-band, we used the photometry from 
the 2MASS catalogue (Cutri et al. \cite{cutri03}) and for the $I$-band 
data we used the photometry from the DENIS catalogue (DENIS Consortium 2003, 
Epchtein et al. \cite{ed1994}).
We have tranformed the $I$-band data from DENIS into the Cousins system, 
using $I_{\rm C}-i_{\rm DENIS} = $ 0.030 $\pm$ 0.04, obtained by comparison of the photometry of M
and L dwarfs from Legget et al. (\cite{leg00}, \cite{leg01}) and DENIS.
We performed an independent calibration for each field and filter. 
Estimates for the completeness magnitudes of the $I$-, $J$- and $K_{\rm
s}$-band images are 21.6 $\pm$ 0.5, 19.2 $\pm$ 0.5, 17.1 $\pm$ 0.5,
respectively. 
Limiting magnitudes of the $I$-, $J$- and $K_{\rm s}$-band images are 22.5 $\pm$
0.5, 20.0 $\pm$ 0.5, 17.8 $\pm$ 0.5, respectively. For the $K_{\rm s}$ images obtained 
with {\sc liris}, the completeness and limiting magnitudes are 18.5 $\pm$ 0.5 and 19.5 $\pm$ 0.5, respectively. 
We adopted as completeness magnitude the value at which the histogram of 
detected objects per magnitude reaches a maximum, and as limiting magnitude the
value at which it drops below half of the maximum (these are roughly equivalent to
a 10$\sigma$ and 3$\sigma$ detection limits, respectively).

\section{Detectability of nearby companions (statistical analysis)}
\label{estadistica}

If we consider that the number of objects with possible companions is well
represented by a Poisson distribution, the probability of having no companions
is given by: 

\begin{equation}
P[x\equiv0] = \left[ e^{-\lambda}
\frac{\lambda^x}{x!}\right]_{x\equiv0}=e^{-\lambda}, 
\end{equation}

\noindent where $\lambda$ is the Poisson parameter, $\lambda=np$, being $n$ the
number of events (8 observed Upper Scorpius members) and $p$, the real
fraction of companions.  
We can determine, for a confidence level (C.L.), where $P =$ 1 $-$ C.L.,
that the upper limit to the real fraction of substellar companions is $p = -ln(P)/n $.

In order to know the real fraction of substellar companions at physical
(not projected) distances, it is necessary to correct this result by the
probability, $p'$, that the orbital position of the companion is at a detectable
projected distance. 
This correction factor assumes dramatic importance when the angle of
inclination, $i$, between the plane of the orbit and the plane of the sky is
close to $\pi / 2$ and the semimajor axis is small. 
The projection of a circular orbit on the plane of the sky is an ellipse with
eccentricity $\sin{i}$:

\begin{equation}
\rho (r,\theta,i) = \frac{r}{(1 + \tan^2{i} ~ \sin^2{\theta})^{1/2}},    
\end{equation}

\noindent where $r$ is the orbital radius and $\theta \in$ [0,2$\pi$).
Assuming that all the angular inclinations of the orbit are equally probable and
that orbits are circular, the probability $p'$ is given by:

\begin{equation}
p'(i_0) = \frac{2}{\pi} \,i_0 +
\frac{4}{\pi^2} \int_{i_0}^{\frac{\pi}{2}} di\,\arcsin{ \left(
\frac{\tan{i_0}}{\tan{i}} \right)},    
\end{equation}

\noindent where $i_0 (r) =\arccos{(r_0/r)}$ and $r_0$
is the projected distance from which a particular companion can be detected. 
In this case, 

\begin{equation}
\lambda(r) = \sum_1^{\rm n}{p'_j(r) ~ p},
\end{equation}

\noindent where $p'_{j}(r)$ is the probability that a companion orbiting at a
distance $r$ is at a detectable separation from the object $j$.  
In a similar way than before, we can determine, for a confidence level and 
a physical separation r, the upper limit to the real fraction 
of substellar companions.

\end{document}